# Complex Network in Solar Features

Somayeh Taran[*]

Department of Theoretical Physics and Astrophysics, Faculty of Physics, University of Tabriz, P.O.Box 51666–16471, Tabriz, Iran
October 13, 2024

# 1  Abstract

This paper is an overview of studying the solar features in a complex network approach. First, we introduce the structural features of complex networks and important network parameters. Applying the detrended fluctuation and rescaled range analysis and nodes degree power-law distributions confirmed the non-randomness of the solar features complex networks. Using the HEALPix pixelization and considering all parts of the solar surface under the same conditions, as well as applying centrality parameters (the nodes with the highest connectivity, closeness, betweenness, and Pagerank) showed that the active areas on the solar surface were correctly identified and were consistent with observations. A review of the complex structure of the solar proton flux and active regions also showed that in these networks, the average clustering coefficient and Page-rank parameters are suitable criteria to use in event prediction methods. The complex network of sunspots has also shown that sunspots and sunspot groups are formed through complex nonlinear dynamics.

# 2  Introduction

In space weather studies, the Sun is of great importance as the considerable influencing factor in space weather conditions near the Earth (Beer et al., 2000;

[*]E-mail: taran@znu.ac.ir

Echer et al., 2005; Pulkkinen, 2007; Schrijver et al., 2015). Various phenomena have been identified in the Sun, the most energetic of which include coronal mass ejections and flares, which have attracted more attention in the scientists' studies (Cane et al., 2002; Tajfirouze and Safari, 2011; Alipour and Safari, 2015; Honarbakhsh et al., 2016; Raboonik et al., 2016; Chen, 2017; Filippov, 2019).

Since solar observational data (for example recorded by Geostationary Operational Environmental Satellite (GOES) and Solar Dynamics Observatory (SDO)) are available in a huge volume, scientists use statistical investigation to predict the occurrence of high-energy phenomena. Raboonik et al. (2016) used Zernike moments of magnetogram data in machine learning method and showed whether an arbitrary active region had the potential to produce at least one large flare. Thereby, they predicted the occurrence of most energetic flares (M and X) 48 hours before they occurred. Using a coronal magnetic field model derived from observational data (by the Hinode satellite), Muhamad et al. (2017) performed 3D magnetohydrodynamic simulations to gain a conception of the flare trigger mechanism. Alipour et al. (2019) applied the support vector machine classifier to Zernike moments of active region image (in ultraviolet and extreme ultraviolet wavelength and magnetograms). They predicted flare events within 4 to 10 days before it.

Dynamic models of flux transfer in the Sun and the existence of different dipole and quadrupole states in solar magnetic periods indicate a complex and asymmetric pattern of the solar dynamo and consequent magnetic field (Schüssler and Cameron, 2018; Russell et al., 2019).

Using the complex network approach, it is possible to study the relationship of a large amount of scattered and unstructured data (e.g., mathematics, physics, computer science, biology, and sociology) and provide models to predict their behavior (Albert and Barabási, 2002; Amaral, 2022). To give an example, the use of the complex network approach in the field of power networks resulted in important insights and increased the strength and stability of the underlying phenomena effective in the power network (Edelman et al., 2018). In investigating earthquakes, it is tough to consider all factors affecting the movement of faults and their mathematical formulation. Therefore, applying a complex network approach is suggested to study it (Rezaei et al., 2017)

The complex network method considering the dynamic properties of the Sun is a suitable proposal for studying nonlinear time series in the Solar features.

Various solar phenomena have been studied in a complex network approach. We discuss the details of these studies as follows: in Section 3, we outline the main characteristics of a complex network. In Section 4 the solar flare complex network models are presented in chronological order. The solar proton flux complex network is describe in Section 6. Section 7 gives an overview of solar active region complex network. Section 8 is also an overview of the complex network of sunspots. The conclusion is given in Section 9.

## 3 Complex network characteristics

The complex networks of different phenomena have different natures, but their structure is similar because they are all governed by the same principles and similar mathematics can be used to study them. Graph theory is used to describe the behavior of complex networks (Strogatz, 2001; Van Steen, 2010; Estrada, 2012). Generally, in mathematical representation, a network is described by an adjacency matrix that provides complete information on the number of nodes and their connections. The number of nodes determines the dimensions of the adjacency matrix. Adjacency matrix can be directed or undirected as well as weighted or unweighted. The components of the weighted matrix contain values between 0 and 1, and the non-weighted matrix contains values of 0 or 1.
Various characteristics of a complex network are measurable, and we mention some of the most important ones.

- **Degree of node distribution**: The degree distribution displays the connectivity of each node in a graph and provides the probability of finding a node with a certain degree k. The power-law behavior of the probability density function (PDF) and the value of the power-law index ($P(k) \propto k^{-\alpha}$) express the small-world and scale-free properties of a complex network (Barabási and Albert, 1999; Myers, 2003).

- **Average shortest path length**: This parameter, which measures the average shortest path length between each pair of vertices in a network and is also called the network's diameter is calculated as follows,

$$\bar{l}s = \frac{1}{N(N-1)} \sum_{i \neq j, i,j=1}^{N} d_{i,j}, \quad (1)$$

  that $d_{i,j}$ and $N$ are the shortest path length between $i$ and $j$ nodes and the total number of nodes, respectively. $d_{i,j}$ can be calculated in directed and undirected graphs by Floyd Warshall's algorithm (Floyd, 1962).

- **Average clustering coefficient**: The tendency of a node to form a cluster with its neighbors is one of the features investigated in the network structure. The clustering coefficient indicates if node $i$ is connected to $j$, and $j$ connects to $m$, with a high probability, that node $i$ connects to $m$. In an undirected and unweighted graph, the local clustering coefficient of node $i$ in a symmetric matrix $A$ is defined as

$$C_i = \frac{1}{k_i(k_i - 1)} \sum_{j,m=1}^{N} A_{ij} A_{jm} A_{mi}, \tag{2}$$

that $A_{ij}, A_{jm}, A_{mi}$ are the number of edges jointing node $i$ to node $j$, $j$ to $m$, and $m$ to $i$, respectively. The average clustering coefficient of the network is given by

$$\bar{C} = \frac{1}{N} \sum_{i=1}^{N} C_i. \tag{3}$$

- **Closeness Centrality**: It is the closest distance between two nodes, which is topologically expressed by using the shortest path length between two nodes with the following equation,

$$C_c(i) = \frac{N - 1}{\sum_{i \neq j} l_{ij}}. \tag{4}$$

- **Betweenness Centrality**: It is defined as the number of shortest paths between two pairs of vertices that lie along the desired edge and sum over all pairs of vertices. For a node $i$, betweenness is defined as

$$C_B(i) = \frac{1}{(N-1)(N-2)} \sum_{m \neq i, j,\ i \neq j} \frac{R_{mj}(i)}{R_{mj}}, \tag{5}$$

where $R_{mj}$ is the number of shortest paths between $m$ and $j$ that passes through $i$.

- **PageRank centrality**: It is a measure that provides the quality and quantity of a node's connections by considering the degree of a node and the degrees of its neighbors.

The average clustering coefficient of an equivalent random network to a real complex network with ($N$) nodes and ($E$) edges is as follows (Fronczak et al., 2004; Boccaletti et al., 2006),

$$\bar{C}_{\text{rand}} = \frac{E}{N^2}. \tag{6}$$

# 4 Solar flares in complex networks approach

Solar flares as high-energy and fast processes are crucial and attractive fields for solar physicists. Within the following, we analyze the various solar flare models studied in complex networks.

## 4.1 Gheibi et al. (2017) model for solar flare complex networks

Gheibi et al. (2017) proposed the first solar flare complex network model. They used 14,395 solar flare information from January 1, 2006, to July 21, 2016, from http://www.lmsal.com/solarsoft/latest_events_archive.html. The main issue in the first approach of building a solar flare complex network was whether the behavior of solar flares showed the complexity of their network or not. In this manner, it is necessary to answer the question of whether the time evolution of the number of flares occurrence has a long correlation with the time series itself. The detrended fluctuation analysis (DFA) determines the self-affinity of a signal. If the value of the Hurst exponent (H) in DFA ranges in (0.5, 1), the time series has a long-term correlation (Mandelbrot, 1975; Alipour and Safari, 2015). Hurst exponent of solar flare time series obtained 0.86. This key suggests that solar flares are governed by critical self-organization, confirming the complexity of the network.

To construct the solar flare complex network, they divided the solar surface into $n \times n$ cells with equal area. If each cell contains flares, it is considered a node, and the flare occurrence time sequence is considered the edge connecting the nodes (connection pattern provided by Abe and Suzuki (2006) in earthquake complex network). Figure 1 shows the divided cells and flare locations. Since the cell length is not uniform across solar latitudes, the tiny cell size results in the network's more accurate working. Considering the spatial distribution of flares is mostly around the solar equator, about 45% of the cells are devoid of flares and are not considered nodes in the network.

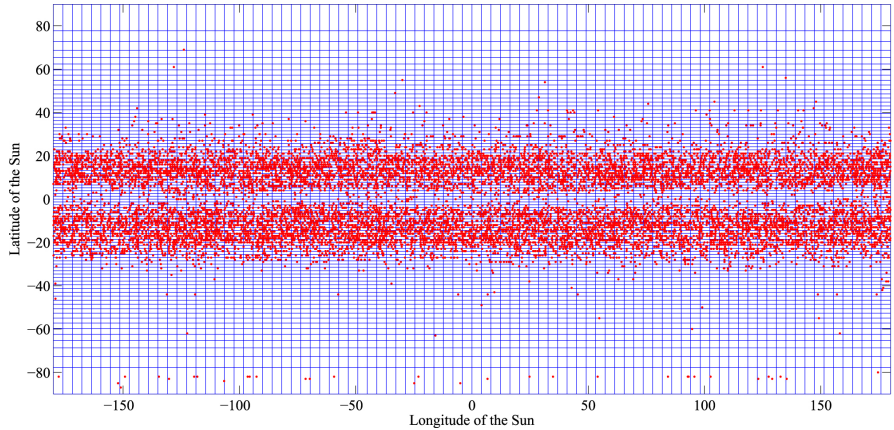

Figure 1: The sun's surface (longitude and latitude) is divided into 88 × 88 cells with equal area. The locations of the flares are placed in the cells, and the empty cells are removed from the flare network analysis. About 45% of the cells are considered nodes in the flare network. © AAS. Reproduced with permission Gheibi et al. (2017).

The degree of node distribution is also an exclusive feature in networks that estimates the complexity. In random networks, the degree difference between different nodes is in the order of the average degrees of the nodes. Consequently, in a random network, the distribution of node degree complies with a Poisson distribution,

$$P(k) = \frac{\exp^{-\lambda} \lambda^{-k}}{k!}, \quad (7)$$

where $k$ and $\lambda$ are the degree of node and a positive constant, respectively. However, in a nonrandom network, due to the presence of nodes with a very high degree, the node degree distribution follows a power-law behavior and denotes the network is scale-free,

$$P(k) \sim k^{-\lambda}, \quad (8)$$

where $\gamma$ is the degree exponent with a positive value. The power-law behavior is the most well-known feature of scale-free systems (Dorogovtsev and Mendes, 2001; Wu, 2013). The power-law function expresses the largest number of nodes have few connections to each other, while a small number of the important nodes have high connections. In this network, the high-degree nodes were introduced as hubs and important nodes. The possibility of flaring in network hub positions is greater than in other nodes. In the smaller size of the network, there is a lack of information or background noise effect, and the network parameters do not show power-law behavior. Consequently, Gheibi et al. (2017) used Aschwanden (2015) step keys and fitted thresholded power-law functions to the probability distribution function of the degree of nodes,

$$P(k) \sim (k + k_0)^{-\lambda}. \tag{9}$$

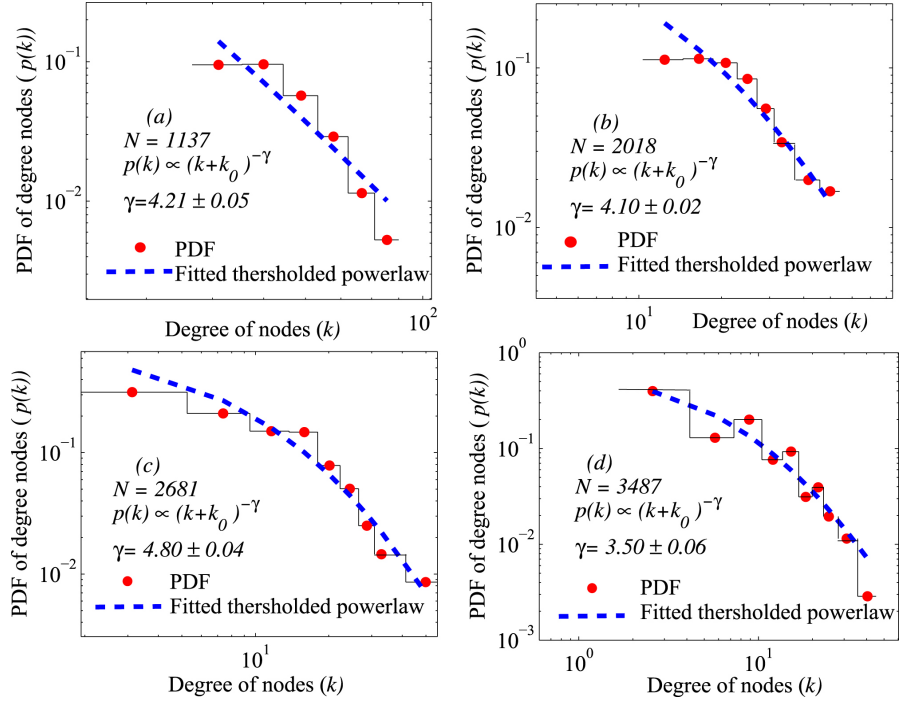

Figure 2: PDFs for the degree distributions of the flare networks displayed in a log-log scale for the network sizes: (a) 1137, (b) 2018, (c) 2681, and (d) 3487. The degree exponents for the power-law fits for different network sizes are found to be greater than three. © AAS. Reproduced with permission Gheibi et al. (2017).

According to Figure 2, by growing the network the degree exponent of power-law distribution remains more than 3. It indicates that solar networks of any size build a small-world network (Cohen et al., 2003). Their results showed that in 15% of the sun's surface, the probability of high-energy flares (M and X) is almost twice as high as in other regions. Although hubs may be located near each other, they

## 4.2 Najafi et al. (2020) model for solar flare complex networks

Najafi et al. (2020) also studied the flares complex network based on the modified model of Gheibi et al. (2017). In the modified network, they tried to reduce the amount of overlap in mapping the solar surface into equal cells. Using the Abe and Suzuki (2006) model and the visibility graph theory (Lacasa et al., 2008), they built the flare complex network modified model. In the visibility graph any event as the node (at time $t_a$, peak energy $F_a$) connects to the node ($t_b$, $F_b$) via the visibility condition if an arbitrary node ($t_c$, $F_c$) with $t_a < t_c < t_b$ obeys

$$F_c < F_b + (F_a - F_b)\frac{t_b - t_c}{t_b - t_a}. \tag{10}$$

Figure 3 shows a small part of the Gheibi et al. (2017) complex network (up panel) and the modified network (down panel) of solar flares. The number of nodes is the same in both models, but the number of edges increased in the modified model. Since each cell is considered a node in the network, the existence of a loop (the connection of the cell with itself) is not far from expected. In the modified model, the loops are removed in the adjacency matrix of the network.

Tsallis and Brigatti (2004) proposed that a q-exponential function can describe distributions with power-law behavior at their tails, which describes the complementary cumulative distribution of events with sizes larger than $S$,

$$P_S = (1 - (1 - q)\beta S)^{\frac{1}{1-q}}, \tag{11}$$

where $P_S$, $q$, and $\beta$ are Cumulative distributions for events with sizes larger than S, nonextensive parameter, and events rate, respectively. $q$ displays the internal properties of the system and describes a wide range of complex phenomena (Tsallis et al., 2003; Lotfi and Darooneh, 2013). Najafi et al. (2020) used the two thresholded power-law and the q-exponential functions to model the node degree distribution and applied the chi-square test with genetic algorithm and MLE in the Bayesian framework to determine the threshold limit in the power-law function.

By constructing a matrix, they examined the characteristics of the network in resolutions ranging from $n = 44$ to $n = 88$. Figure 4 presents the solar flare time series for the selected data set. According to the visibility graph conditions, larger events connect to small and medium-sized events which increases the degrees of main flare events. The number of edges reaches saturation with increasing resolution.

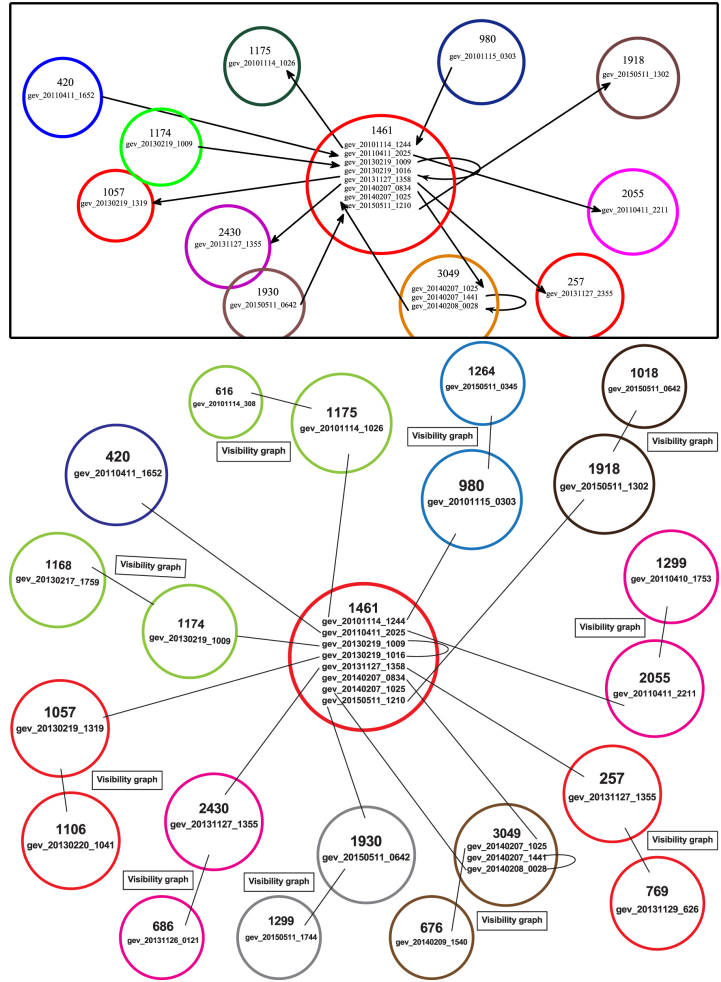

Figure 3: A small part of the Gheibi et al. (2017) solar flare network (up panel) and the modified network (down panel) are shown. © AAS. Reproduced with permission Gheibi et al. (2017) and Najafi et al. (2020).

To prove the non-randomness of the network, they compared the solar flare clustering coefficient with the clustering coefficient of the equivalent random network. The sizeable difference between them confirmed the non-randomness of the network (Figure 5a). Also, by increasing the network resolution, the ratio of the regular and random network clustering coefficients increased (Figure 5b). Due to the increase of connections in the modified network, the clustering coefficient

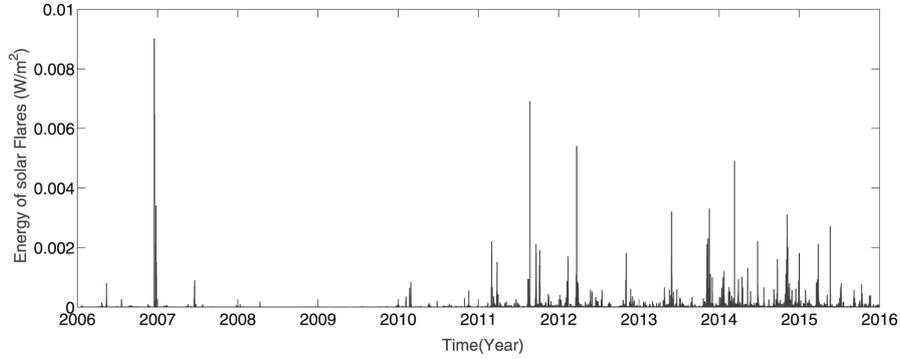

Figure 4: Energy time series of 14,395 solar flares between January 1, 2006, and July 21, 2016. © AAS. Reproduced with permission (Najafi et al., 2020).

increased compared to the Gheibi et al. (2017) network.

Examining the behavior of the clustering coefficient of the nodes with increasing network size and for different resolutions showed that the modified network follows the power-law function like the previous network and the scale-free characteristic of the solar flare network was confirmed in both networks.

In the modified network, the number of nodes remains constant compared to the previous model, but with the increase of links, it is possible to check the compliance of the flare network with some empirical laws. Ōmori (1894) studied the earthquake data and stated Omori's law that the frequency of events decays after a main event. Since the energy frequency of solar flares and earthquakes follows a power-law distribution, they investigated the modified flare network following Omori's law, and the frequency of large flares after the main flare decreases.

## 4.3 Solar flare ultraviolet signatures

Lotfi et al. (2020) have applied an unsupervised network-based method to detect the location and timing of solar flares. They selected three regions (flaring and non-flaring active region, and the quiet-Sun regions) on the solar surface at solar ultraviolet (1600 Å) emission to analyze the network parameters. The data were selected from the SDO/AIA images with 24 seconds time cadence. The selected data covered the period before the start of flaring until after it. Table 1 in Lotfi et al. (2020) presents the selected area information at channel 1600 Å.

Since the sun has a differential rotation, for the correct processing of the data, this effect must be removed from the images and all of them must be rotated relative to a reference image (Tajik et al., 2023). Figure 6 shows the selected regions in colored boxes. Black dashed boxes are regions that contain flare-active regions (a, b, d, and e) and non-flare active regions (c). Regions with different colors are chosen to cover flaring regions, regions close to the flare event location, and quiet-sun regions away from active regions. The selected box dimensions were 100 x 100 pixels. To reduce noise and improve calculation speed, they examined the average of each 5 x 5 pixels in a 20 x 20 bins box. Each bin in a 20 x 20 box was converted into a light curve. The 400 optical curves had the role of nodes in the network. The time interval chosen to increase the network size is 50 frames, and all regions overlap simultaneously in 40 frames. Using Pearson correlation, the nodes in consecutive images were linked as the network's edge. If the correlation level was more than 0.5, the nodes connected, else there was no connection between them. They stated that introducing small correlations (≤ 0.5) increased connections and caused a loss of significant changes in network parameters, and large correlations (> 0.5) made the network more sparse.

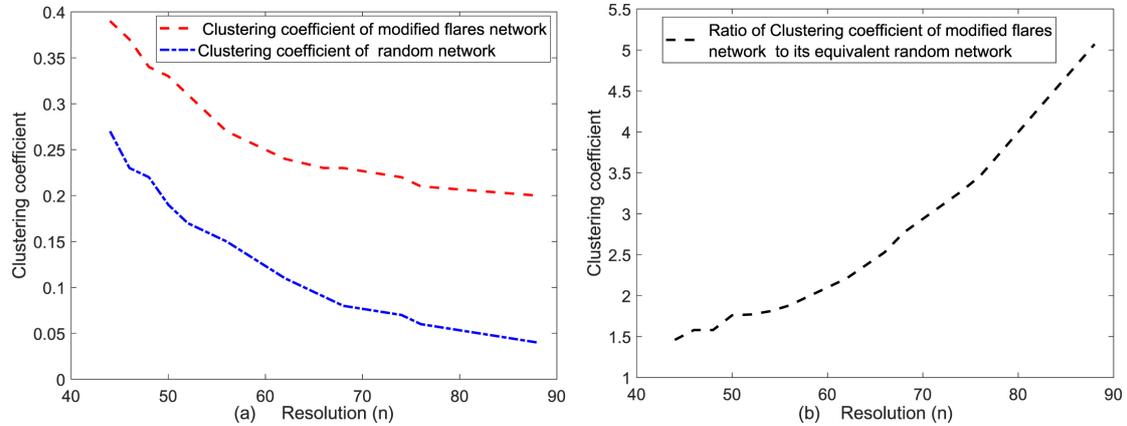

Figure 5: Clustering coefficient of solar flare modified grid and its equivalent random grid in terms of resolution (a). The $\frac{C}{C_{rand}}$ ratio, that increases with resolution. © AAS. Reproduced with permission Najafi et al. (2020).

In presenting the results, all the network parameters were normalized to the equivalent random network parameters. The results of calculating the network parameters showed that at the active flaring areas, the normalized clustering coefficient ($C/C_{rand}$) value decreases suddenly during flaring time. However, in other active non-flaring regions, the $C/C_{rand}$ value fluctuates slightly around the average value (Figure 7).

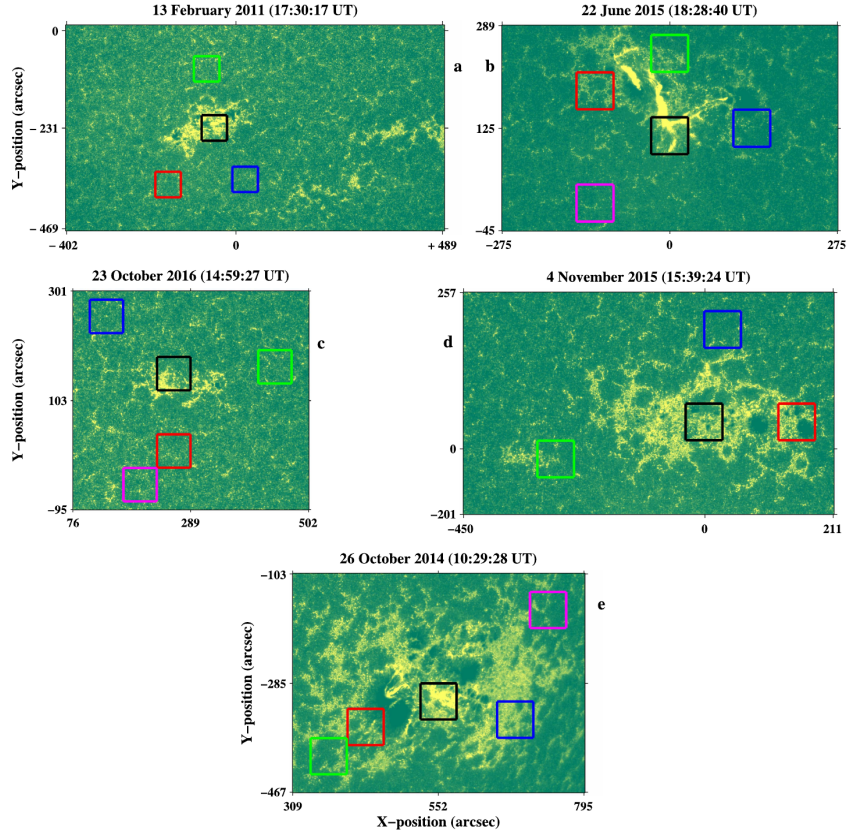

Figure 6: De-rotated cutout images of SDO/AIA at 1600 Å. Boxes with black contours are areas with about 60 × 60 arc secs including flaring ARs (panels (a), (b), (d), and (e)) and non-flaring AR (panel (c)). Reprinted from Lotfi et al. (2020), with the permission of AIP Publishing.

Comparing the normalized clustering coefficient of different regions revealed that the $C/C_{rand}$ value of the non-flaring active regions is $> 6$, and $\leq 6$ in the flaring area. As a result, the low value of the $C/C_{rand}$ compared to the average value can show signs of flare occurrence earlier than the flaring time.

As a measure of activity in complex networks, they analyzed the PageRank for 30 minutes before the flaring start time, during the flare occurrence, and 30 minutes after that. The results indicated that the page rank magnitude increases during the flare event in the flaring areas and their nearby regions, but the quiet sun areas

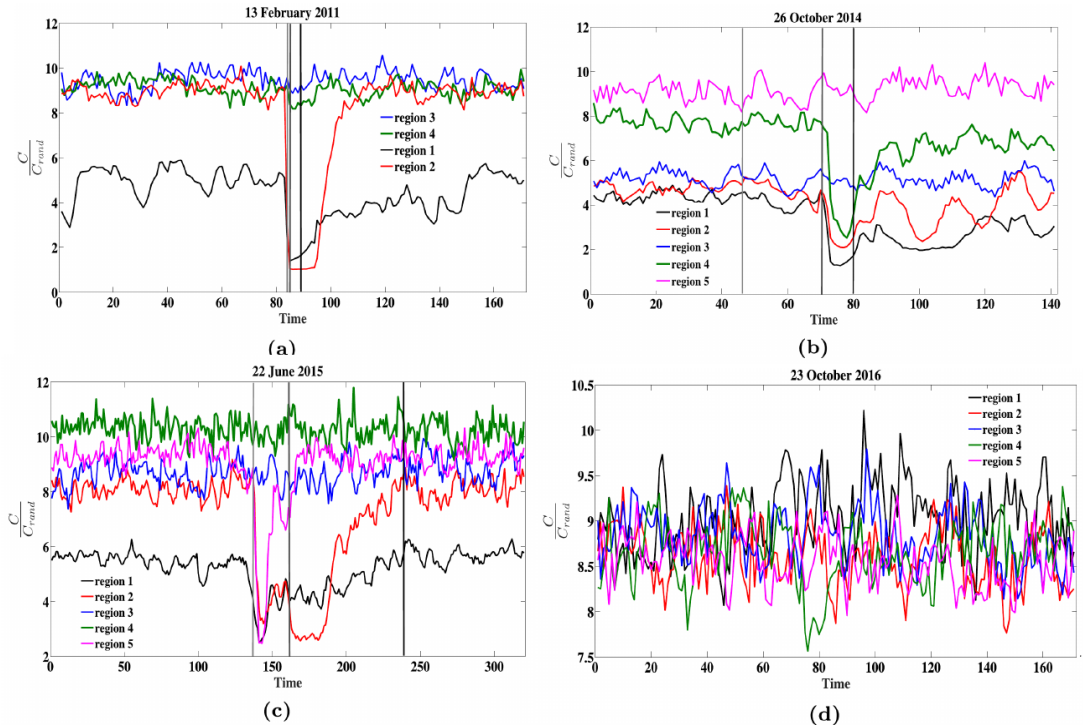

Figure 7: The time series of the normalized clustering coefficients of the network belonging to different regions of the Sun. The color of each region corresponds to the coloring of the boxes in Figure 2, which cover different solar regions. Reprinted from Lotfi et al. (2020), with the permission of AIP Publishing.

were not affected by the flare event (see Figure 7 in Lotfi et al. (2020)).

## 5 Taran et al. (2022) model for solar flare complex networks

Taran et al. (2022) could eliminate the error caused by the superposition of cells in the previous divisions by using a new method of dividing the solar surface into perfectly identical and uniform areas. Using the HEALPix method of Gorski et al. (2005), they divided the surface of the solar sphere into rhombic-spherical cells so that the center of cells is in a similar solar latitude. In the HEALPix pixelization, the grid resolution is expressed by the resolution parameter $N_{\text{side}} = 1, 2, 4, 8, \ldots$ and the total pixel number is $N_{pix} = 12N_{side}$. In various resolution parameters, each rhombic-spherical cell area is omega. In various resolution parameters, each rhombic-spherical cell area is omega. Figure 9 shows dividing the surface of a sphere for different resolutions in the HEALPix method.

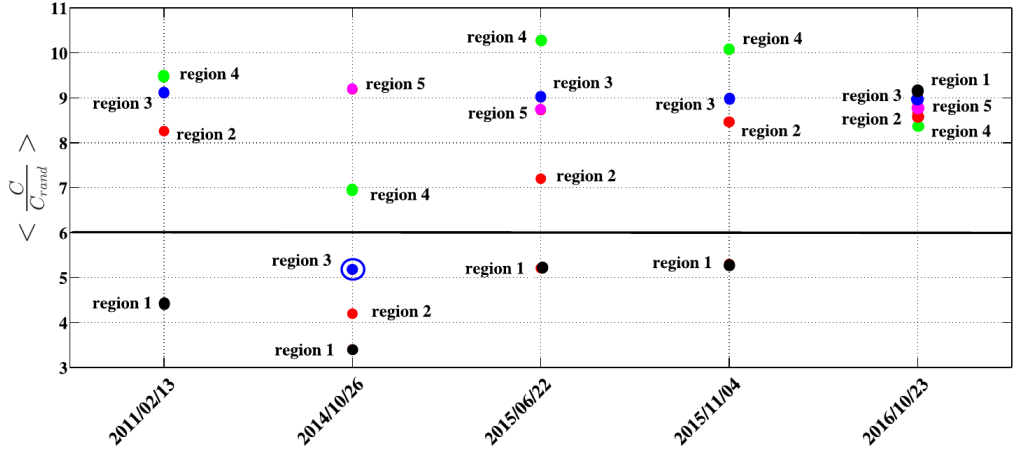

Figure 8: The average value of $C/C_{rand}$ was obtained from the time series of Figure 7 for different regions and dates. Regions with mean values of 6 belong to active regions including large flares. Reprinted frLotfi et al. (2020), with the permission of AIP Publishing.

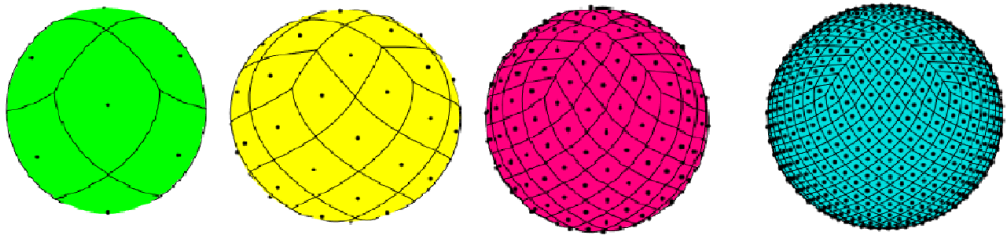

Figure 9: From left to right, HEALPix spherical rhombic structures for $N_{side}$ = 1, 2, 4, and 8, respectively. Dots represent the center of each pixel. Image reproduced with permission from Taran et al. (2022), copyright by Elsevier.

Taran et al. (2022) used 8766 solar flare data with fluxes more than $5 \times 10^{-6} W/m^2$ (C, M and X classes) from 2008 December to 2019 December taken from the Hinode Flare Catalogue (https://xrt.cfa.harvard.edu/flare_catalog/). In this work, the focus was on investigating the activity of the solar hemispheres in the solar cycle 24. The share of the northern and southern hemispheres in this cycle was 4033 and 4725 flares, respectively.

Figure 10 shows the distribution of solar flare occurrence location for two resolution parameters 2 (a) and 8 (b) on HEALPix unite radius sphere

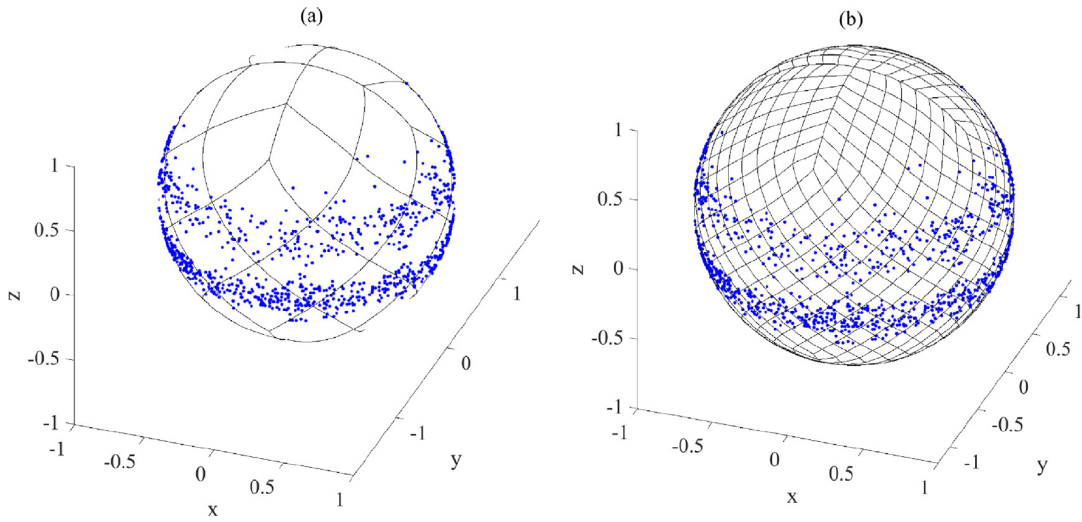

Figure 10: The view of the map of solar flare locations is represented on the HEALPix sphere for $N_{side}$ = 2 (a) and $N_{side}$ = 8 (b). Image reproduced with permission from Taran et al. (2022), copyright by Elsevier.

They used the HEALPix and visibility graph approach to construct the solar flare network. Each pixel is considered a node, and an edge connects two cells according to the visibility graph algorithm. The flares that occur inside each cell are considered a node. They construct an unweighted and undirected graph by ignoring loops as two successive earthquakes connect in the same cell and replacing multiple edges between the cells with one edge. Considering 25% of the nodes with the highest value determined the four criteria (degree of nodes, closeness, betweenness, and Pagerank) and then selecting the common nodes among these four criteria as hubs, Taran et al. (2022) proposed a new method to identify hubs in the network. The results of calculating the network parameters were displayed in the Mollweide projection. Figure 11 shows the results of identifying hubs for different resolutions ($N_{side}$ = 1, 2, 4, 8, 16 and 32). According to the color bar, the color of each cell shows the magnitude of the network parameter. This figure recognizes the flaring active areas. The Mollweide projection provides better and more interpretable results by increasing the resolution in the HEALPix division. The results show that the number of hubs in the Northern Hemisphere is much higher than in the Southern Hemisphere. But in high resolutions, super hubs were detected in the southern hemisphere. They proposed that flares occur in these areas with a considerable probability.

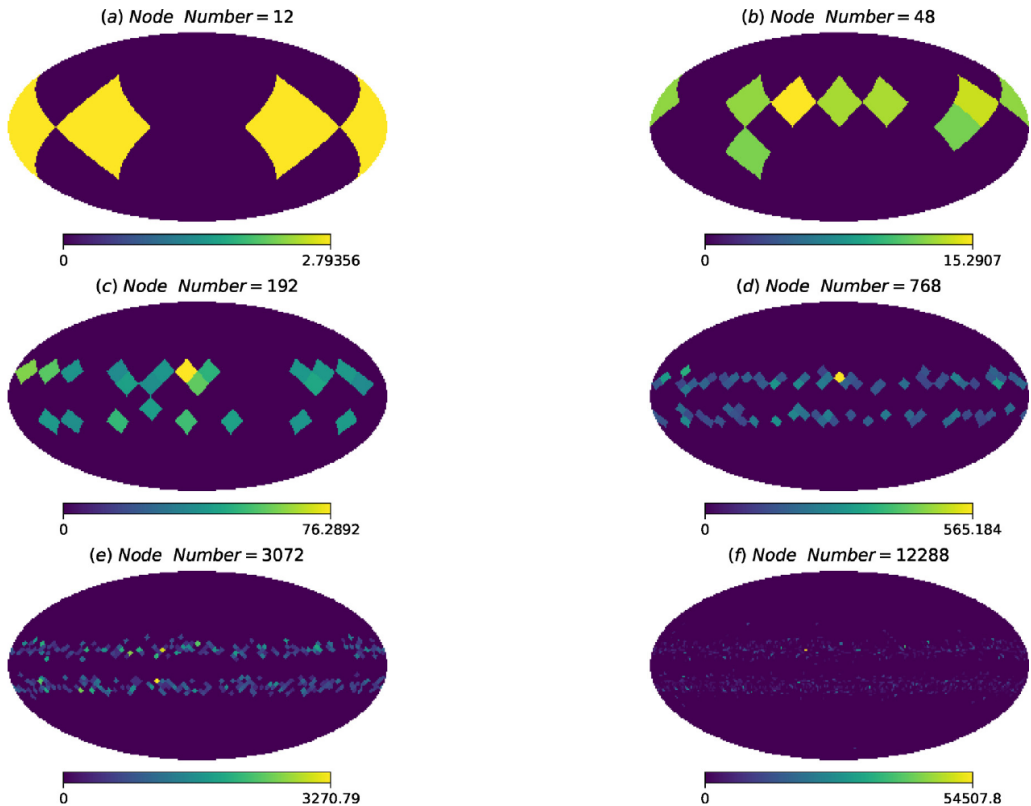

Figure 11: Detected hubs in the northern and southern hemispheres for HEALPix cells along with the number of nodes at different resolutions Nside = 1 (a), 2 (b), 4 (c), 8 (d), 16 (e), and 32. Image reproduced with permission from Taran et al. (2022), copyright by Elsevier.

# 6   Solar proton flux complex network

Mohammadi et al. (2021) studied the properties of the solar proton flux network and its correlation with solar flares by using the GOES 13 and 15 data sets from 14 April 2010 to 9 May 2018. The studies were carried out in 6 energy intervals ($> 1, > 5, > 10, > 30, > 50$, and $> 100$ MeV) of solar protons with data time cadence 5 minutes. Each interval has 14729 data. To know the solar flares occurring time, they used the Interactive Multi-Instrument Databaseflare solar flare catalog. The considered flares included flares along with coronal mass ejections and fast coronal mass ejections (more than 500 km/s). In the construction of the complex network, they used the optimized visibility graph algorithm. Lan et al. (2015) raized the time complexity of visibility graph $O(n^2)$ to $O(n \log n)$ in optimized alghorithm. $n$ is the number of nodes in the network. The proton flux and its time are considered as a node in visibility graph theory. A representation of the visibility graph of protons at energies greater than 30 MeV is shown in Figure 12.

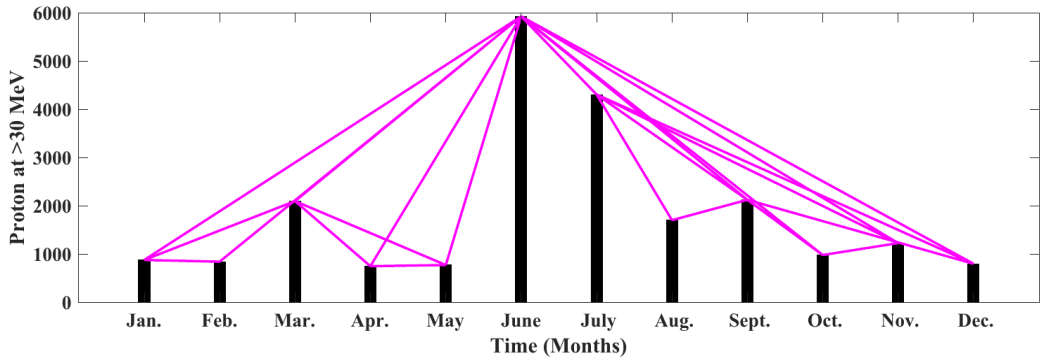

Figure 12: An example of the visibility diagram for proton flux (at ¿30 MeV) is provided. The horizontal axis represents the months of January to December and the vertical axis shows the monthly integrated proton flux during 2011. Referring to the visibility condition, the nodes are connected by straight lines. Image reproduced with permission from Mohammadi et al. (2021), copyright by John Wiley & Sons.

Applying rescaled range analysis showed the long-range dependencies in solar proton flux time series. The Hurst exponents value for all 6 energy ranges was 0.7 and confirmed the complexity of the solar proton flux network. Investigations showed that for the network dimensions less than 1001, the adjacency matrix of the network forms a complete graph. Consequently, Mohammadi et al. (2021) studied the network for sizes greater than 1001 to identify the key features.

Comparison of the ratio of solar proton flux network clustering coefficient to equivalent random network coefficient ($C/C_{\text{rand}}$) with increasing size of the network (for example, for energies greater than 30 MeV in Table 1 in Mohammadi et al. (2021)) confirms the non-randomness of the network.

According to Figure 13, the length scale of the solar proton flux network versus the network size for all considered energy intervals in the semi-logarithmic space shows the linearity of this relationship that confirms the network small-world characteristic (Newman and Watts, 1999). It shows the small-world property of the networks for three channels ($> 1, > 5, and > 10 Mev$) with threshold 1001, and for the high energy channels ($> 30, > 50, and > 100 Mev$) appears for sizes more than 2001.

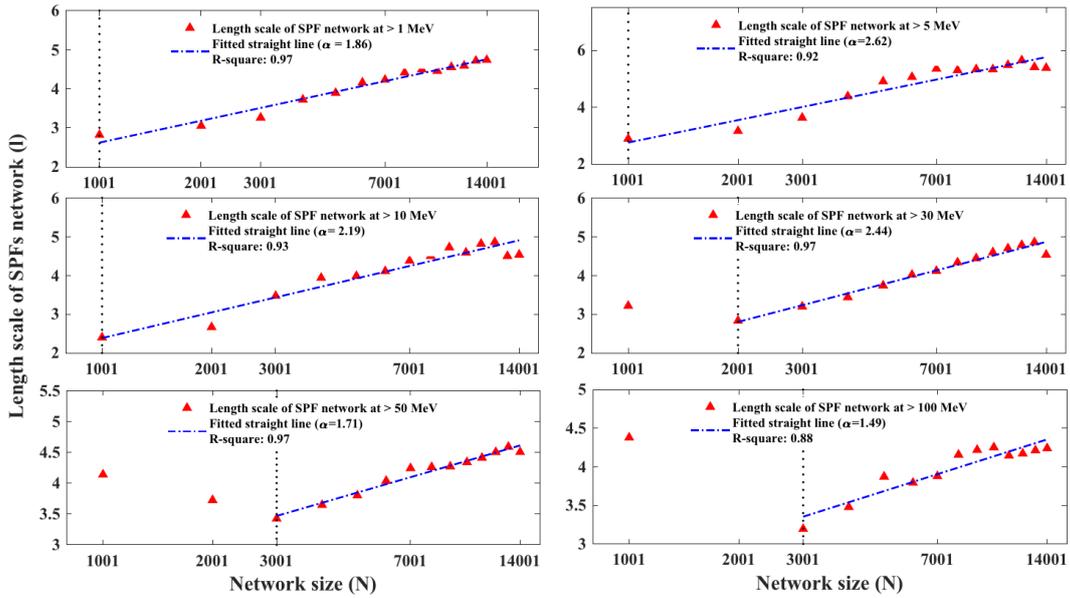

Figure 13: The solar proton flux grid length scale (red triangle) is displayed in terms of the network size (N). The fitted line applies to the relation $l \propto \log_{10} N$. Vertical dashed lines indicate thresholds. For each fit, the R-squared goodness-of-fit measure is shown. Image reproduced with permission from Mohammadi et al. (2021), copyright by John Wiley & Sons.

An interesting part of Mohammadi et al. (2021) article is investigating the correlation between solar flare events and proton networks. The average clustering coefficient and PageRank of the complex network were selected for correlation analysis. Figure 14 shows the jumping of average clustering coefficients on the large flare (X and M) occurring time (for example for energy ¿ 50 MeV).

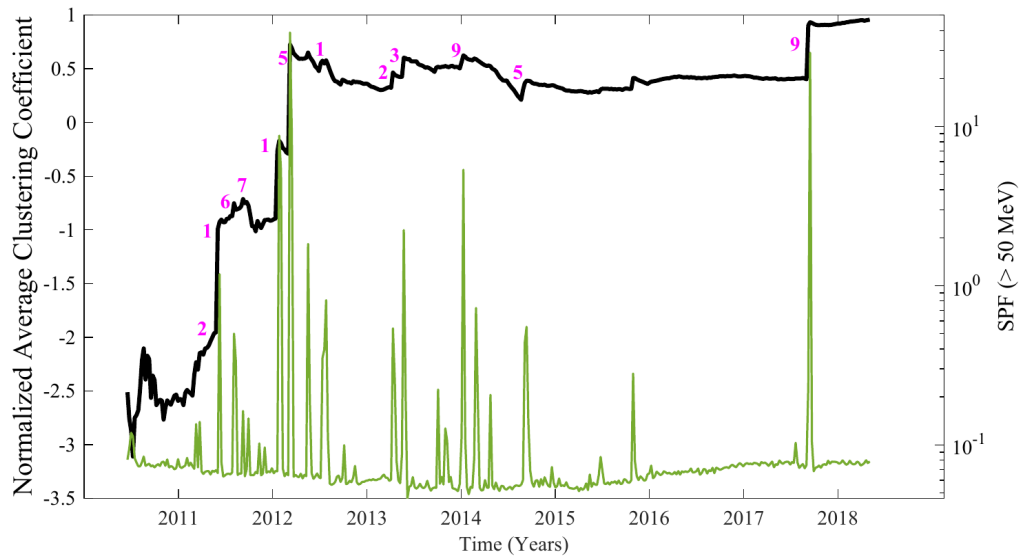

Figure 14: The weekly solar proton flux time series (green line) and average clustering coefficient (black line) are shown for solar proton flux growth (at more than 50 mV). The time of occurrence and number (within 7 days) of X-class and M-class flares (pink numbers) are shown. Image reproduced with permission from Mohammadi et al. (2021), copyright by John Wiley & Sons.

In space weather, it is essential to identify the events that produce energetic solar particles. Many studies have been done in this field, and a significant correlation between solar proton flux and flares was found in the long term (Belov et al., 2005; Gopalswamy et al., 2008; Cane et al., 2010; Dierckxsens et al., 2015). Using Extreme Ultraviolet Imaging Telescope (EIT) images, Cliver (2005) showed the arrival of a wavefront near the time that started the activity in the solar western limb and injected the first solar energetic particles toward the Earth. Alberti et al. (2017) also showed that flares occurring in western solar longitudes develop more energetic particles in space than flares that arise in other regions. Mohammadi et al. (2021) found that the network parameters of solar energetic particles (Pagerank) have a higher correlation with the flares associated with the CME than the time series of the particles themselves. However, the correlation magnitude is significantly high for the flares that emit in western solar longitudes (see Tables 2 and 3 and Figure 1 therein).

# 7 Solar active regions complex network

Daei et al. (2017) studied various features of the solar active regions complex network.
Each solar active region is recorded in the solar monitor (`www.solarmonitor.org`) with a NOAA number which allows tracking of the active region in successive images of the Sun. They studied active regions from January 1, 1999, to April 14, 2017. By tracking the NOAA numbers in the solar monitor daily tables, we can obtain the location and lifetime of each active. Referring to the Goes flare catalog, selected active regions include energetic flares (X, M, and C).
Figure 15 shows the position of solar longitude and latitude of the active regions in the $0 \leq \theta \leq 180$, $0 \leq \phi \leq 360$ after removing the solar differential rotation effect.

Applying the DFA and rescaled range analysis on the active region's time series, the hurst exponent value was 0.8 and 0.94, respectively. These values show the autocorrelation of the active regions' time series and the formation of a self-organized criticality (SOC) system. The SOC is an important method in describing the complex nature of nonlinearity and degrees of freedom in many physical and astrophysical phenomena (Wang et al., 2014; Aschwanden et al., 2016). Daei et al. (2017) investigated the SOC behavior of solar active region systems in a complex network framework.
For constructing the active region complex network, based on the Gheibi et al. (2017) method, they divided the solar surface into square cells and considered each cell including the active region a node in the network graph. To reduce the complexity of the calculations, the temporal variation of the solar longitude and latitude of the active regions were ignored and the first place of appearance of each active region was considered as its fixed location in the network. If an active region appears during the lifetime of another one, the corresponding cells are connected and form the edges of the network graph. Figure 16 shows a small part of the active region network consisting of 6 nodes, 15 edges, and 4 loops. According to the explained algorithm, for example, an active region with NOAA number 10081

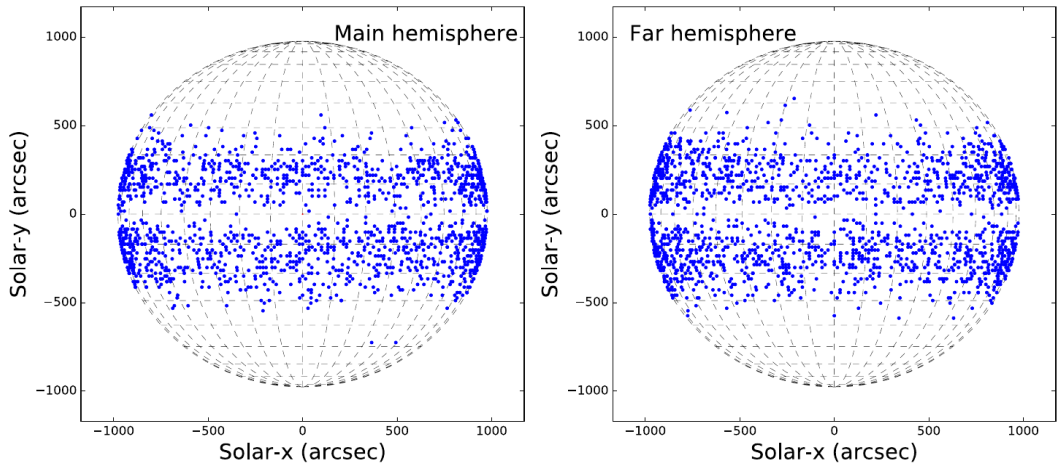

Figure 15: Positions of 4227 active regions (from 1999 January 1 to 2017 April 14) at their first occurrence times, rotated concerning the first active region (NOAA 8419) coordinates. All positions' longitudes and latitudes are restricted to 0°–360° and 0°–180°, respectively, and are mapped to the surface of a sphere. © AAS. Reproduced with permission Daei et al. (2017).

was formed during the lifetime of an active region with NOAA number 10063.

Investigating the active region network parameters provided the following results. The difference between the values and behavior of the clustering coefficient of the active region network and the equivalent random network is also a confirmation of the non-randomness of the active region network. The length scale of the network had a small value and its behavior regarding the network size was logarithmic that confirmed the active regions network is in the category of small-world networks (Figure 17).

Daei et al. (2017) showed that the degree distribution of the active regions network follows the power-law. This behavior indicates the network is scale-free and includes active hubs. Also, Observations confirm the result and show that energetic phenomena such as flares or coronal mass ejections occur in just some of the active areas not all of them (Nitta et al., 2012; Toriumi and Wang, 2019; Iglesias et al., 2020). Daei et al. (2017) observations showed that the number of flare occurrences increased in cells identified as hubs, and on average, the number of high-energy flares occurring in hubs was at least twice as high as in other cells (Figure18).
As a result, in the methods of flare prediction, active regions are good candidates for the location of flare occurrence. However, providing predictive methods requires more extensive studies.

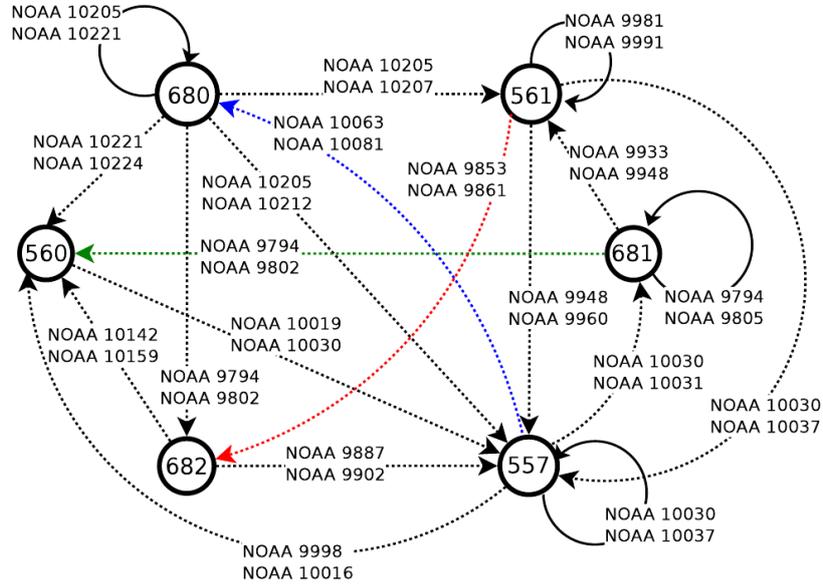

Figure 16: A small part of the active region network containing six nodes is shown. Each arrow represents a connection edge between two nodes. Each loop shows the occurrence of two simultaneous active regions in one node. © AAS. Reproduced with permission Daei et al. (2017).

## 8  Sunspots Complex Networks

Sunspots are a phenomenon that clearly shows the complexity of the solar magnetic field (McIntosh et al., 2014; Song and Zhang, 2016). Studying the sunspots time series generally reveals the periodicity of solar activities, although the amplitude of sunspot activity differs in various periods. Sunspots with different numbers and sizes appear on the solar surface, and observations show that large spots are divided into tiny spots, which is a sign of the self-organization of the sunspot system (Consolini et al., 2009; Shapoval et al., 2018). Mohammadi Gouneh et al. (2023) investigated the complex network of sunspots. They used 15,816 sunspot data from February 31, 1922, to February 31, 2016, recorded in https://www.sidc.be/silso/datafiles. Data information in-cludes sunspots' number, size, solar latitude, and longitude of event location.
Applying the DFA method to the sunspots' time series showed that the value of the Hurst index is between 0.5 and 1, which is a confirmation of the existence of autocorrelation in the time series of sunspots and the non-randomness of the sunspots' complex network.

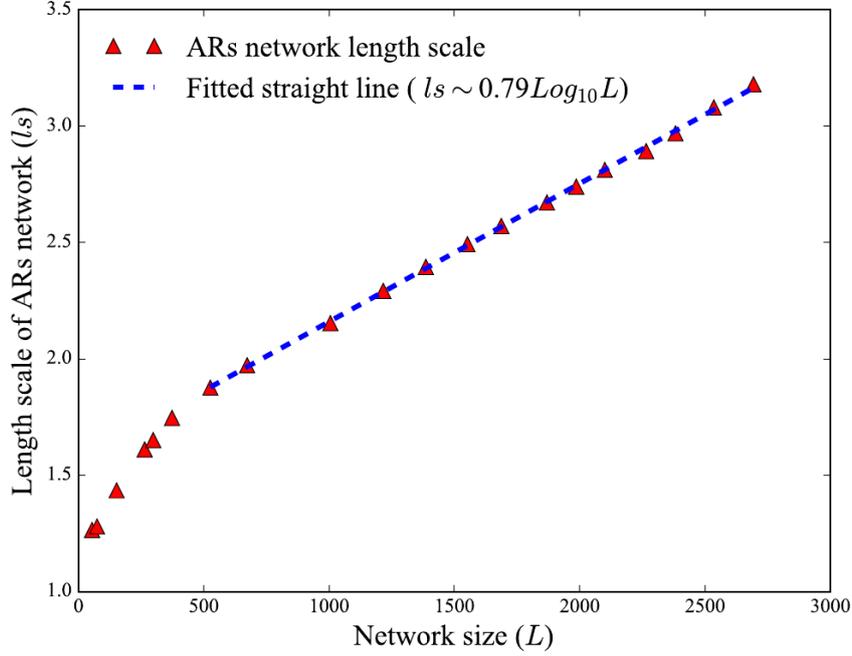

Figure 17: The length scale of the active region network (triangle) in terms of the network size (L) and a fitted straight line as $l_s \sim 0.79 \log_{10} L$. © AAS. Reproduced with permission Daei et al. (2017).

In the network construction, Mohammadi Gouneh et al. (2023) choose the number of spots at any time as nodes and connect these nodes according to the visibility graph. Figure 19 shows a visibility graph diagram of the sunspot network.

They constructed an undirected and unweighted graph and showed that the degree distribution of the sunspots network nodes follows the thresholded power-law distribution function.Consequently the sunspot network is a scale-free network. Figure 20 shows the distribution of node degrees and the fitted power-law function for different network sizes.

The clustering coefficient of the sunspots network, in addition to being fundamen-tally different from the clustering coefficient of the equivalent random network, regardless of the various values of the network size, occupies an almost constant value in the range of 0.65 to 0.67. Referring to Table 2 in Mohammadi Gouneh et al. (2023), the maximum network diameter is 10 and the characteristic path length is 4.5. It means that nodes are connected by at most 10 links to any other arbitrary node and on average, the nodes are connected by 4.5 links. As a result, the spots are in related groups, which is also confirmed in the observations. This group behavior can be caused by the complex dynamo of the Sun. The magnetic orientation of sunspots in the northern and southern hemispheres of the sun and the placement of pairs of sunspots with opposite magnetic polarities in the east-west direction in solar latitudes close to each other is a sign of the assortativity of the sunspot network. Figure 21 shows the correlation function of the degree of network nodes in terms of the degree of nodes for different network sizes in the log display.

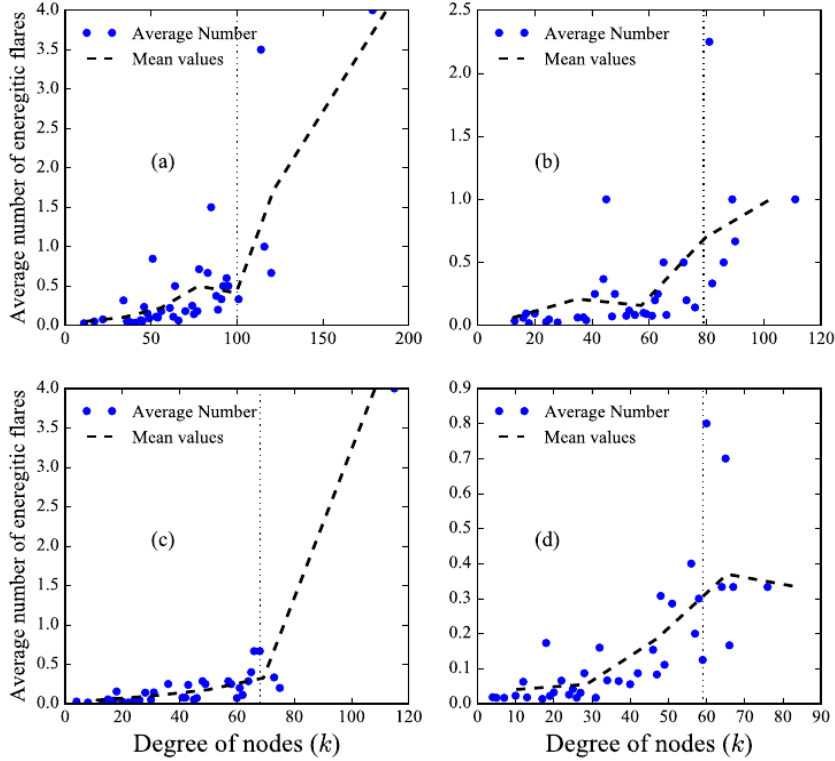

Figure 18: Average number of flares (circles) and average connected values of non-uniform binning (dashed line) observed at the position of the nodes of the active region network versus the degree of the nodes for the network size (L) (a) 1552, (b) 1986, (c) 2382 and (d) 2693. Nodes with high degrees (hubs) are presented to the right of the dotted lines. © AAS. Reproduced with permission Daei et al. (2017).

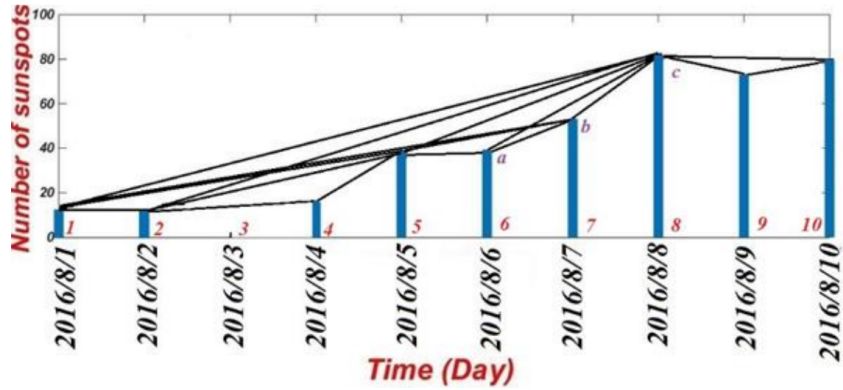

Figure 19: The visibility graph is shown schematically for 10 days. A counter from 1 to 10 is assigned to each day. Image reproduced with permission from Mohammadi Gouneh et al. (2023).

According to the Figure 21, the sunspot complex network for different network sizes is assortative, dis-assortative, or neutral. But to give a general conclusion based on the largest size of the network, the network is assortative. This result can be changed by increasing the number of sunspots. Coherence implies that sunspots with different polarizations can be related to each other.

# 9 Conclusion

Scientists have proposed different algorithms for building complex networks of natural phenomena (Boccaletti et al., 2006; Mitchell, 2006; Havlin et al., 2012; Zeng et al., 2017). However, to provide an algorithm that can reveal all the complex behaviors of a phenomenon requires the examination of different patterns during an evolutionary process. The complex network approach in solar phenomena study is a practical method to investigate the solar events' nonlinear and collective interaction (Aschwanden et al., 2016). In this article, I attempt to create a general perspective by reviewing the articles on solar features' complex networks. It helps

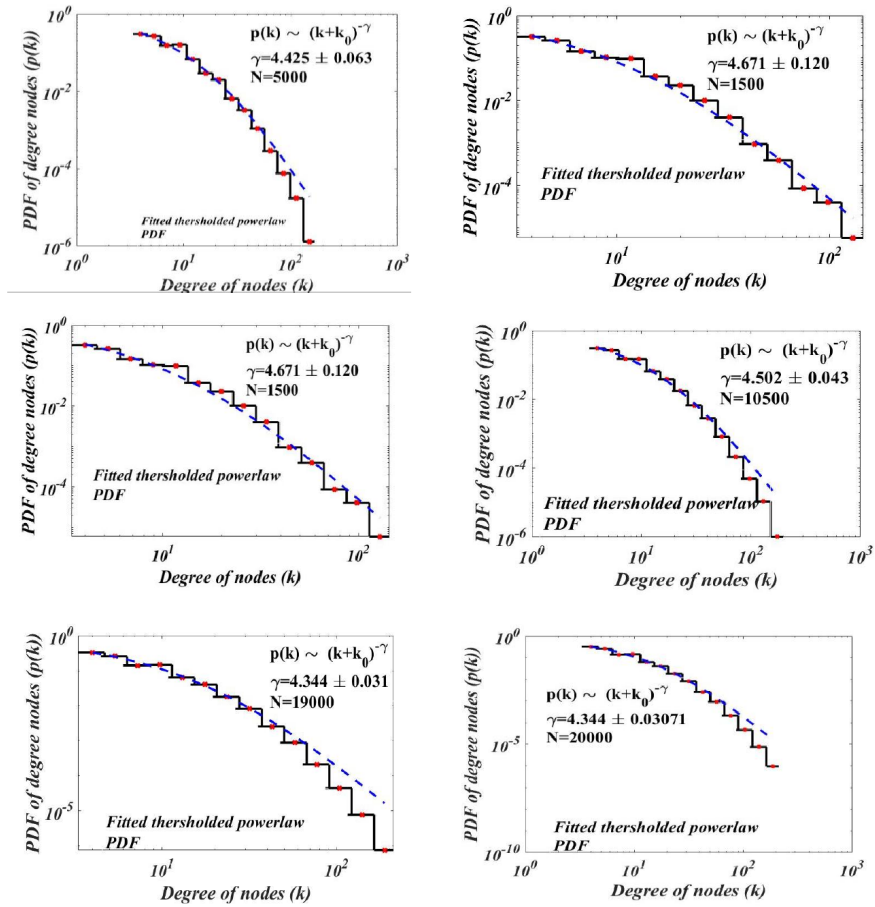

Figure 20: Showing the distribution of degrees of the complex network time series of sunspots in full logarithmic scale (square) and fitting the thresholded power-law function (dashed line) with network sizes from 1500 to 20000. Image reproduced with permission from Mohammadi Gouneh et al. (2023).

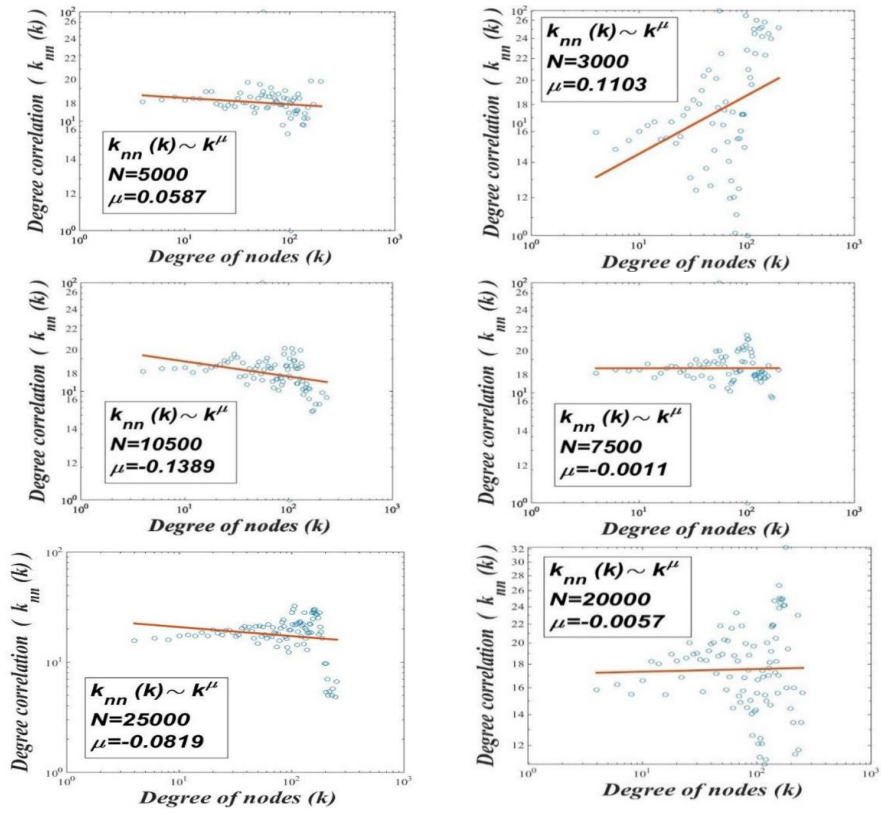

Figure 21: The degree correlation function of the sunspot complex network time series (circle) and the fitted linear function in the full logarithmic for the network size of 3000 to 25000. Image reproduced with permission from Mohammadi Gouneh et al. (2023).

us to study more detailed and comprehensive patterns of solar event mechanisms in complex networks.

Gheibi et al. (2017), in the first model of the solar flares complex network, confirmed that the flares network is not random, and the interaction of the occurring flares is formed according to a specific mechanism. This confirmation was enough to examine the flare network by providing a more detailed algorithm. Najafi et al. (2020) modified the Gheibi et al. (2017) model and increased the number of interactions compared to the previous algorithm. They showed that solar flare networks similar to earthquakes, follow Omori's natural law. In both models, the solar spherical surface was divided into quadrangular shapes on a flat plane. This type of division will be associated with errors in areas near the solar poles, and the equality of cells is not respected in the entire solar sphere. However, most solar flares occur around the equator and this will cover the solar surface pixelization error. Taran et al. (2022) used the HEALPix pixelization and divided the solar surface into rhombic-spherical cells with the same shape and area. In the accurate division of the spherical surface into the same areas and shape as well as the simultaneous use of centrality parameters (the nodes with the highest connectivity, closeness, betweenness, and Pagerank), it is guaranteed that the flare active areas (hubs in complex network) are identified correctly and with high accuracy.

Considering the extreme influence of the earth and space weather from the solar features mechanisms, predicting the behavior of these phenomena is one of the critical approaches in solar studies (Nandy, 2021; Georgoulis et al., 2024). Lotfi and Darooneh (2013), in studying solar ultraviolet images, and Mohammadi et al. (2021) in investigating solar protons and their correlation with flares, showed that the parameter of clustering coefficient and page rank in complex networks are suitable quantities for predicting the event of flares. Investigating the sunspot's complex network parameters showed the existence of scale-free and small-world characteristics of the time series network of sunspots.